\documentclass[letterpaper, 10 pt, conference]{ieeeconf}  

\IEEEoverridecommandlockouts                              
\usepackage{graphicx}      
\usepackage{amsmath}
\usepackage{amssymb}
\usepackage{mathrsfs}  
\usepackage{algorithm} 
\usepackage{hyperref}
\usepackage{arydshln}
\usepackage{algpseudocode} 
\newcommand{\ip}[2]{\ensuremath\left\langle{#1}, {#2}\right\rangle}
\newcommand{\dom}{\ensuremath\mathrm{dom}\,}

\title{\LARGE \bf
An Operator-Theoretic Framework to Simulate Neuromorphic Circuits
}
\author{Amir Shahhosseini, Thomas Chaffey and Rodolphe Sepulchre
\thanks{A. Shahhosseini is with KU Leuven, Department of Electrical Engineering (STADIUS), KasteelPark Arenberg, 10, B-3001 Leuven, Belgium, {\tt\small amir.shahhosseini@kuleuven.be}. T. Chaffey is with the University of Cambridge, Department of Engineering, Trumpington Street, CB2 1PZ, {\tt\small tlc37@cam.ac.uk}. R. Sepulchre is with both the KU Leuven, Department of Electrical Engineering (STADIUS), KasteelPark Arenberg, 10, B-3001 Leuven, Belgium, and the University of Cambridge, Department of Engineering, Trumpington Street, CB2 1PZ, {\tt\small rodolphe.sepulchre@kuleuven.be}.}
\thanks{The research leading to these results has received funding from the European Research Council under the Advanced ERC Grant Agreement SpikyControl n.101054323. The work of T. Chaffey was supported by Pembroke College, Cambridge.}%
}

\begin{document}

\maketitle
\thispagestyle{empty}
\pagestyle{empty}

\begin{abstract}

Splitting algorithms are well-established in convex optimization and are designed to solve large-scale problems. Using such algorithms to simulate the behavior of nonlinear circuit networks provides scalable methods for the simulation and design of neuromorphic systems. For circuits made of linear capacitors and inductors with nonlinear resistive elements, we propose a splitting that breaks the network into its LTI lossless component and its static resistive component. This splitting has both physical interpretability and algorithmic tractability and allows for separate calculations in the time domain and in the frequency domain. To demonstrate the scalability of this approach, a network made from one hundred neurons modeled by the FitzHugh-Nagumo circuit with all-to-all diffusive coupling is simulated. 

\end{abstract}

\section{Introduction}

Neuromorphic engineering promises a new technological paradigm for computing machines. Neuromorphic technologies are designed to be more energy efficient than their digital counterparts \cite{mead1990neuromorphic} and handle ill-conditioned inputs significantly better \cite{furber2016large}. In recent years, spiking neural networks \cite{lobo2020spiking}, neuromorphic computing devices \cite{young2019review}, and neuromorphic sensors \cite{vanarse2016review} have drawn the attention of scientists in various disciplines. This interest creates new demand in the simulation, analysis, and design of large-scale electrical circuits that are at the heart of any neuromorphic system. 

The common methods of simulating spiking neurons rely on the {\textit{numerical integration}} of nonlinear differential equations \cite{sejnowski1988computational}. This approach enabled the study and simulation of simple neurons and contributed to the development of computational neuroscience. Nevertheless, numerical integration methods are known to struggle with spiking behaviors due to the stiffness of the governing equations. Additionally, such methods do not scale up adequately with an increase in the size of the network. It is also impractical to analyze the robustness of the neural networks against the variability and uncertainty of the components. Finally, numerical integration methods cannot include and exploit prior knowledge about the system's behavior within their structure: they are not {\textit{behavior-informed}} solvers. This motivates and necessitates exploring alternative methods for the efficient simulation and analysis of large-scale electrical circuits that represent neuromorphic systems.

This paper takes inspiration from passive linear time-invariant (LTI) circuit theory in developing an alternative. There, studying the network is a computationally tractable task, and passivity provides insight into the physics of the problem \cite{kalman1963lyapunov}. Motivated to extend these properties to networks with nonlinear resistors, Minty proposed the concept of {\textit{monotonicity}} \cite{minty1960monotone}. In the meantime, Rockafellar's works on convex analysis have placed monotonicity at the core of scalable convex optimization algorithms \cite{bauschke2017correction}. Thus, like passivity in the case of LTI networks, monotonicity is the concept that reconciles the physics of nonlinear circuits with algorithmic tractability. It is a key concept that permits the development of new efficient methods for the simulation of large-scale nonlinear networks.

To this end, the recent papers \cite{chaffey2023monotone,chaffey2023circuit} explore monotone circuits, and formulate their dynamics as a monotone {\textit{zero inclusion problem}}. This is then broken down using splitting algorithms and its solution is obtained by solving a {\textit{fixed-point iteration}} (FPI) \cite{ryu2022large}. Splitting algorithms are central to these studies and their distributed nature grants the computational tractability of large-scale optimization methods. In fact, by allowing each circuit component to be handled independently, the algorithms make complex computations efficient and manageable. The results of these studies confirm that methods of large-scale convex optimization can be used instead of numerical integration for the {\textit{scalable}} simulation of monotone circuits and networks. 

Contrary to monotone circuits, models of spiking neurons contain a mixture of monotone and anti-monotone elements that represent the mixed feedback nature of such systems \cite{sepulchre2022spiking}. The recent work  \cite{das2022oscillations} explores the framework of convex-concave programming \cite{shen2016disciplined} to solve mixed monotone systems. Nevertheless, no attention is given to the splitting itself. In fact, the question of efficient, scalable, and behavior-informed simulation and analysis of spiking systems remains open.

Similar to how monotonicity captures the physics of circuit elements, splitting should capture the circuit topology. In addition to providing physical intuition into the structure of the network, a {\textit{proper splitting}} also makes simulation and computations significantly faster. This paper exploits a splitting of the circuit into its lossless elements and its dissipative nonlinear resistors. With this splitting, the zero inclusion problem is broken into two subproblems, one being an LTI lossless operator (representing inductors, capacitors, and interconnections with no dissipation) and the other a mixed-monotone operator (the difference of two monotone operators representing the nonlinear resistors). 

This paper solves nonlinear spiking systems not by using numerical integration methods but by utilizing operator-theoretic methods and efficient FPIs. It also provides an energy-based splitting that comes with both physical interpretability and clear computational advantages. We call this approach \emph{time-frequency splitting}. This method also allows users to exploit their prior knowledge of the spiking network's behavior in the solver. This shortens the simulation time by orders of magnitude.

The remainder of this paper is organized as follows. Section 2 formulates the problem, and describes the energy-based splitting and its connection with the energy-based methods of the literature. Section 3 presents an algorithm that can solve this zero-inclusion problem. It also explains how using both the time and frequency domains allows the exploitation of the LTI structure of the lossless part and makes the computations efficient. Section 4 demonstrates the presented algorithm by simulating the FitzHugh-Nagumo model \cite{izhikevich2006fitzhugh} of a spiking neuron and, extends this idea to a network of 100 heterogeneous neurons with diffusive coupling. 

\section{Energy-based Splitting of Nonlinear Circuits}

In this section, the energy-based splitting of RLC networks with static nonlinear resistors is introduced. Furthermore, the connection between this representation and existing energy-based representations of nonlinear circuits is highlighted.

\subsection{Preliminaries}

We begin by introducing some necessary preliminaries. $\mathcal{L}^2_\mathbb{T}$ is the Hilbert space of square-integrable signals over the time axis $\mathbb{T}$, equipped with the inner product

\begin{equation}
    \ip{u}{y} = \int_{\mathbb{T}} u^\top(t)y(t)dt < \infty
\end{equation}

We use the shorthand notation $\mathcal{L}^2$ to denote $\mathcal{L}^2_{[0, \infty)}$, and $\mathcal{L}^2_T$ to denote $\mathcal{L}^2_{[0, T]}$. The latter may be associated with $T$-periodic signals, restricted to a single period. A trajectory $v(t)$ is $T$-periodic if $v(t+T) = v(t)$ for any $t$. The Hilbert space of $T$-periodic signals is of interest since we mostly deal with periodic behavior. The space $l^2 _T$ is the discrete-time counterpart of $\mathcal{L}^2 _T$. 

\textbf{Definition 1.} An operator $\operatorname{A}$ on the space $\mathcal{L}^2$ is a set-valued
mapping $\operatorname{A}: \mathcal{L}^2 \rightrightarrows \mathcal{L}^2
$. The {\textit{graph}} of the operator $\operatorname{A}$ is defined as
\begin{equation}
    \text{Gra} \operatorname{A} = \{(u,y) | y \in \operatorname{A}(u)\} \subseteq \mathcal{L}^2 \times \mathcal{L}^2.
\end{equation}

\textbf{Definition 2.} An operator $\operatorname{A}: \mathcal{L}^2 \rightrightarrows \mathcal{L}^2
$ is {\textit{monotone}} if 
\begin{equation}
    \ip{u_1 - u_2}{y_1 - y_2}  \geq 0
\end{equation}
for all $u_1, u_2 \in \dom {\operatorname{A}}$ and $y_1, y_2$ are the corresponding outputs. The operator $\operatorname{A}$ is said to be \emph{maximal monotone} if its graph is not properly contained in the graph of any other monotone operator.

\textbf{Definition 3.} An operator $\operatorname{A}: \mathcal{L}^2 \rightrightarrows \mathcal{L}^2
$ is {\textit{n-cyclic monotone}} if, for every $(u_1,u_2,...,u_{n+1}) \in \mathcal{L}^2$ and every $(y_1,y_2,...,y_n) \in \mathcal{L}^2$ where $u_1 = u_{n+1}$ and $y_i = \operatorname{A}(u_i)$ 
\begin{equation}
    \sum_{i=1}^n\left\langle y_{i+1}-y_i , u_i\right\rangle \leq 0
\end{equation}
If $\operatorname{A}$ is $n$-cyclically monotone for every integer $n \geq 2$, then $\operatorname{A}$ is {\textit{cyclically
monotone}}. An operator $\operatorname{A}$ is {\textit{maximal cyclic monotone}} if it is maximal monotone and cyclic monotone.

\textbf{Theorem 1.} (Rockafellar’s theorem \cite{bauschke2017correction})
An operator $\operatorname{A}: \mathcal{H} \rightrightarrows \mathcal{H}
$ is {\textit{maximal
cyclic monotone}} if and only if it is the subgradient of a closed, convex, and proper function from $\mathcal{H}$ to $(-\infty,\infty]$.

\textbf{Definition 4.} Given an operator $\operatorname{A}: \mathcal{L}^2 \rightrightarrows \mathcal{L}^2
$ the \emph{resolvent} $\operatorname{J}_{\alpha \operatorname{A}}: \mathcal{L}^2 \rightrightarrows \mathcal{L}^2
$ is defined as
\begin{equation}
    \operatorname{J}_{\alpha \operatorname{A}} = (I + \alpha \operatorname{A})^{-1}
\end{equation}
where $\alpha$ is a scalar and $\alpha > 0$.

\subsection{Neuromorphic Circuits and Energy-based Splitting}

\subsubsection{From Neural Networks to Electrical Circuits}

Neuromorphic engineering aims to build electronic circuits that mimic biological nervous systems. Neurons are the building blocks of these nervous systems, and for decades, their working mechanisms have been the focus of studies within neuroscience  \cite{squire2012fundamental}. They compute and transfer information through action potentials (spikes) that are nothing more than a sudden discharge of ions. The ion channels that govern the passing of these {\textit{electrical charges}} are best modeled as nonlinear memristive elements. The neuron’s membrane can be modeled using a capacitor. 
This fundamental alignment of the physical language of the neurons with electrical elements motivates the modeling and analysis of neuromorphic systems using electrical circuit theory.

\begin{figure}[h]
    \centering
\includegraphics[scale=0.5]{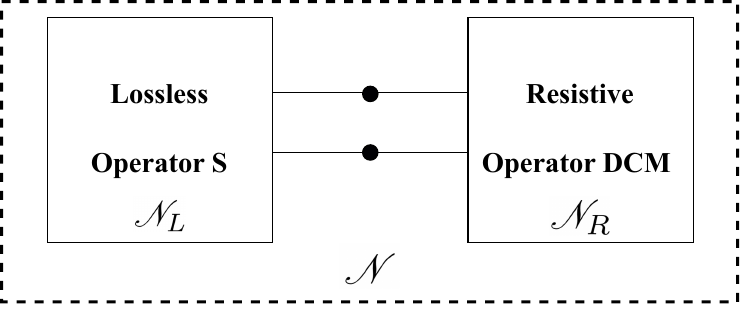}
    \caption{Decomposition of the circuit network $\mathscr{N}$ into a lossless part and a resistive part}
    {\label{fig:decomposition}}
\end{figure}

It is possible to utilize network analysis within circuit theory to describe the dynamics of electrical circuits. However, the representation of the dynamics is not unique. The representation of this paper relies on the energy dissipation properties of the elements and breaks the network down into two subcircuits. The first subcircuit contains all the resistive elements and the second subcircuit contains the rest: inductors, capacitors, and interconnections, which are all LTI lossless elements and have no dissipation. This is concurrent with the {\textit{resistance extraction method}} in circuit theory \cite{desoer1969basic}. Figure \ref{fig:decomposition} illustrates the splitting of the general network $\mathscr{N}$ into a lossless subnetwork $\mathscr{N}_L$ and a resistive subnetwork $\mathscr{N}_R$.

\subsubsection{Energy-based Splitting of Nonlinear Electrical Circuits}

The main idea is to express the behavior of the electrical circuit as a zero-inclusion problem with an operator governing the dynamics. Mathematically speaking, 
\begin{equation}{\label{eq:ZIP}}
    \operatorname{A} \begin{pmatrix}
i(t)\\ 
v(t)
\end{pmatrix} = 0,
\end{equation} 
where $\operatorname{A}$ comes from Kirchhoff's voltage law (KVL), Kirchhoff's current law (KCL), and the physics of the elements. In Eq. (\ref{eq:ZIP}), $i(t)$ and $v(t)$ are trajectories in the signal space $\mathcal{L}^2_{[0,\infty)}$ or in case of periodic behavior, $\mathcal{L}^2_T$. Even for simple circuits, $\operatorname{A}$ can become complicated and it is useful to break it into smaller operators. Splitting the operator into smaller operators helps in two ways. First, if done properly, the smaller operators will bear physical meaning and maintain their connection with the underlying circuit elements. Second, proper splitting of the operator makes the problem computationally easier. This is because the {\textit{resolvent operator}} of the original operator is usually difficult to compute while the smaller operators are simpler to deal with \cite{ryu2022large}. This is the main motivation for using splitting algorithms in large-scale convex optimization algorithms \cite{parikh2014proximal}. Furthermore, it is rational to exploit the structure of the smaller operators to make the computations even faster. To this end, Eq. (\ref{eq:ZIP}) can be rewritten as 
\begin{equation}{\label{eq:S+R}}
    \operatorname{S} \begin{pmatrix}
i(t)\\ 
v(t)
\end{pmatrix} + 
\operatorname{R} \begin{pmatrix}
i(t)\\ 
v(t)
\end{pmatrix} = 0,
\end{equation}
where $\operatorname{R}$ represents the resistive portion of the circuit and $\operatorname{S}$ represents the lossless portion. We call this an ``energy-based splitting’’.

\subsection{Mixed Monotone Description of Nonlinear Resistors}

The interplay of active and dissipative elements is central to the spiking behavior. A monotone (incrementally passive) $\operatorname{R}$ cannot generate such a motion and part of $\operatorname{R}$ must supply energy to the system. This translates to $\operatorname{R}$ being mixed monotone. It is also possible to model $\operatorname{R}$ as the difference of monotone operators. Thus, Eq. (\ref{eq:S+R}) can be reformulated as 
\begin{equation}{\label{eq:split}}
    \operatorname{S} \begin{pmatrix}
i(t)\\ 
v(t)
\end{pmatrix} + 
\operatorname{M_1} \begin{pmatrix}
i(t)\\ 
v(t)
\end{pmatrix} - 
\operatorname{M_2} \begin{pmatrix}
i(t)\\ 
v(t)
\end{pmatrix} = 0,
\end{equation}
where $\operatorname{M_1}$ and $\operatorname{M_2}$ are cyclic monotone operators. The cyclic monotonicity of $\operatorname{M_1}$ and $\operatorname{M_2}$ implies that they are the subgradients of convex functionals, via Rockafellar’s theorem (Theorem 1). The reason for breaking the operator $\operatorname{R}$ further into its (cyclic) monotone components is mainly computational. In fact, most algorithms only work when the operators are monotone, and mixed monotone operators disrupt their functionality and convergence \cite{ryu2022large}. Beyond computational advantages, this further splitting also bears physical meaning. $\operatorname{M_1}$ corresponds to the part of the nonlinear resistor that dissipates energy while $\operatorname{M_2}$ is the part that supplies energy. 

\subsection{Connection with Energy-based Representations}

The splitting of this paper relies on the energy characteristics of the circuit components. In this section, we explore the connection of this representation with other energy-based representations of electrical circuits. For circuits with linear elements, dynamics can be obtained by using the Hamiltonian method. For the case of conservative systems, it is only necessary to obtain the Hamiltonian/Co-Hamiltonian. For linear dissipative elements, Wells introduced a ``power function” (corresponding to Rayleigh’s dissipation function in mechanics) that could be incorporated into the Hamiltonian (or Lagrangian) equations and obtain the dynamics of linear RLC circuits \cite{wells1945power}. 

The extension from linear electrical circuits to nonlinear complicates this energy framework. The nonlinear resistive elements could no longer be expressed in terms of simple dissipation functions. However, this issue was resolved when Millar proposed the idea of content and co-content which were generalizations of the “power functions” \cite{jeltsema2005modeling}. 

In fact, the dynamics of an RLC network with nonlinear resistors can be written as 
\begin{equation}{\label{eq:coHamiltonian}}
    \left\{\begin{matrix}
\frac{d}{dt} \nabla _i {H}^*(i,v) + \nabla _i (v^TNi) = -\nabla _i \mathcal{R}(i,v) \\ 
\frac{d}{dt} \nabla _v {H}^*(i,v) - \nabla _v (v^TNi) = -\nabla _v \mathcal{R}(i,v),
\end{matrix}\right.
\end{equation}
where ${H}^*(i,v)$ is the co-Hamiltonian and is defined as 
\begin{equation}
    {H}^*(i,v) = \mathcal{T}^*(i) + \mathcal{V}^*(v).
\end{equation}
Here, $\mathcal{T}^*(i)$ and $\mathcal{V}^*(v)$ are the inductor’s magnetic co-energy and capacitor’s electric co-energy \cite{jeltsema2009multidomain}. The term $\mathcal{R}(i,v)$ is defined as 
\begin{equation}
    \mathcal{R}(i,v) = \mathcal{D}(i) + \mathcal{D}^*(v)
\end{equation}
where $\mathcal{D}(i)$ and $\mathcal{D}^*(v)$ are the corresponding content and co-content of the resistive elements and describe the dissipated power. The relations of Eq. (\ref{eq:coHamiltonian}) are identical to the zero inclusion problem of Eq. (\ref{eq:S+R}). Here, the left-hand side only represents lossless LTI elements and the interconnection of the elements. Moreover, the right-hand side also describes the resistive part of the network.

\section{Time-Frequency Approach in Mixed Monotone Circuits}

Zero inclusion problems of the form (\ref{eq:S+R}) are commonly solved by reformulating them to FPIs. These methods involve iteratively evaluating the resolvent of the original operator. However, computing the resolvent of complicated operators is an exhaustive task. The issue is also compounded if the operator contains both dynamic elements and static nonlinear elements. To this end, operators are split into simpler components, and the FPIs are reformulated to utilize the resolvents of the simpler components. This approach offers significant computational advantages and simplifies solving involved zero inclusion problems. The interested reader is referred to \cite{ryu2022large} for relevant literature and the importance of monotonicity in this framework. 
 
\subsection{Difference of Monotone Splitting Algorithms}
The monotonicity of the operator in the zero inclusion problem ensures that if the problem has a solution, it can be found systematically using computationally tractable algorithms \cite{ryu2022large}. However, the problem is more complex when the operator is not monotone since it can have no solutions, one solution, or many solutions. Additionally, given different initializations for the FPI algorithm, it is possible to converge to different solutions. 

The Douglas-Rachford splitting algorithm is a powerful splitting algorithm that finds the zero of the sum of two monotone operators. This method was recently extended to the class of difference of monotone operators \cite{chuang2022unified,chaffey2022input} that have the form 
\begin{equation}{\label{eq:DMDR+SPLIT}}
    \operatorname{S} \begin{pmatrix}
x
\end{pmatrix} + 
\operatorname{B} \begin{pmatrix}
x
\end{pmatrix} - 
\operatorname{C} \begin{pmatrix}
x
\end{pmatrix} = 0,
\end{equation}
where $\operatorname{S}$, $\operatorname{B}$ and $\operatorname{C}$ are monotone operators and $x$ is a signal belonging to $\mathcal{L}^2_T$ in our problem. As can be seen, this is identical to the format of Eq. (\ref{eq:split}) and thus, this problem can be solved using this method. The pseudo-code for the difference of monotone Douglas-Rachford (DMDR) algorithm is provided below. Here, $\operatorname{J}_{\alpha \operatorname{A}}$ represents the {\textit{resolvent operator}} of operator $\operatorname{A}$ (Definition 4), and $\alpha$ is the step size. $z$ defines the auxiliary variable that is used within the structure of the algorithm and the superscripts denote the iteration count of the algorithm.
\begin{algorithm}
	\caption{DMDR} 
	\begin{algorithmic}[1]
		\For {$j=1,2,\ldots, max-iteration$} 
                \State $x^{j+1}=\operatorname{J}_{\alpha \operatorname{S}}\left(z^j\right)$
                \State $z^{j+1}=z^j-x^{j+1}+\operatorname{J}_{\alpha \operatorname{B}}\left(2 x^{j+1}-z^j+\alpha \operatorname{C}\left(x^{j+1}\right)\right)$
                \State %
		\EndFor
	\end{algorithmic} 
\end{algorithm}

The approach of this paper is not limited to the DMDR algorithm and can be used with any splitting algorithm that can solve the zero inclusion problem of the form of Eq. (\ref{eq:DMDR+SPLIT}).

\subsection{Time-Frequency Approach}

In previous works \cite{chaffey2023monotone,chaffey2023circuit,das2022oscillations}, the dynamical components were discretized in the time domain, using a backward Euler discretization of the differentiation operator. Computing the resolvent of such matrices in the time domain, which involves inversion (as indicated in Definition 4), creates a dense matrix by inverting a structured and very sparse matrix. This does not exploit the structure of the LTI elements and makes the computations inefficient. 

LTI operators are diagonal in the frequency domain and many tools in control theory, such as transfer function, exploit this property. The matrix representation of LTI operators has a circulant structure and their inversion can be performed efficiently in the frequency domain due to this exact property.

In the method of this paper, the resolvent of the operator dealing with the lossless LTI portion of the system is now computed in the frequency domain. To carry out this step in the frequency domain, let us first write the first step (line 2 of pseudo-code) as
\begin{equation}
\begin{aligned}
(\mathbb{I} + \alpha \operatorname{S})\begin{pmatrix}
i^{j+1}(t)\\ 
v^{j+1}(t)
\end{pmatrix} = \qquad \qquad \text{ \:}\\
(\mathbb{I} + \alpha \begin{bmatrix}
P(\operatorname{D}) & N^*\\ 
-N & Q(\operatorname{D})
\end{bmatrix})\begin{pmatrix}
i^{j+1}(t)\\ 
v^{j+1}(t)
\end{pmatrix} = 
 \begin{pmatrix}
z_1^{j}(t)\\ 
z_2^{j}(t)
\end{pmatrix}
\end{aligned}
\end{equation}
where the signal $x$ in the first step of the DMDR algorithm is broken into $v(t)$ and $i(t)$ to provide a form that is compatible with Eq. (\ref{eq:split}). The lossless operator $\operatorname{S}$ is also expanded to show its structure. $P$ and $Q$ are related to the inductors and capacitors and are functions of the derivative operator. $N$ incorporates the interconnection and is a matrix with elements that can only be $+1$, $-1$, or $0$. 

Upon taking the FFT of the signals and representing the LTI operators in the frequency domain, the operator functions $P$ and $Q$ become diagonal. Mathematically
\begin{equation}
\underset{\text{Frequency domain}}{\underbrace{
(\hat{\mathbb{I}} + \alpha \begin{bmatrix}
\hat{P}(j \omega) & N^*\\ 
-N & \hat{Q}(j \omega)
\end{bmatrix})}}\begin{pmatrix}
I^{j+1}(j \omega)\\ 
V^{j+1}(j \omega)
\end{pmatrix}
 =
 \begin{pmatrix}
Z_1^{j}(j \omega)\\ 
Z_2^{j}(j \omega)
\end{pmatrix}
\end{equation}
where $\hat{P}$ and $\hat{Q}$ are now diagonal matrices (that are in the frequency domain) and thus, new iterations of the signals $I(j \omega)$ and $V(j \omega)$ are obtained in the frequency domain. By taking the iFFT of these signals, the new iteration of the signal $x$, which includes both $i(t)$ and $v(t)$, is obtained in the time domain. For the sake of clarity, the entire process of first step is summarized as
\begin{equation}
\begin{aligned}
iFFT
\begin{pmatrix}
(\mathbb{I} + \alpha \begin{bmatrix}
\hat{P}(j \omega) & N^*\\ 
-N & \hat{Q}(j \omega)
\end{bmatrix})^{-1} FFT
\begin{pmatrix}
z_1^{j}(t)\\ 
z_2^{j}(t)
\end{pmatrix}
\end{pmatrix},
\end{aligned}
\end{equation}
but here, the inversion does not deal with a dense matrix.

The computations related to the second step of the DMDR algorithm (line 3 of the pseudo-code) should be carried out in the time domain since static nonlinear elements are in fact diagonal in the time domain. To compute the resolvent, efficient proximal algorithms such as the guarded Newton method can be used \cite{parikh2014proximal}.

By switching between the time and frequency domain, the LTI structure of the lossless components is exploited and the computational cost of simulating the network is drastically decreased. Upon computing the resolvent through finite-difference discretization and not exploiting the structure, the first step involved a general inversion that has a complexity of $\mathcal{O} (n^3)$ but this is now reduced to an elementwise vector by vector multiplication, and the computations of this step are dominated by the computational cost of FFT and iFFT that is $\mathcal{O} (n\log(n))$. Using Fourier transform to invert circulant matrices is a standard method in numerical linear algebra \cite{davis1979circulant}.

\section{Results: From a Single Neuron to a Heterogeneous Network}

To show the capability of the proposed method, two examples are provided. The first example is the FitzHugh-Nagumo model of the spiking neuron. This example is solved in detail as a tutorial. The second example is a network of 100 neurons with all to all connections. This example serves as a proof of concept for the scalability of this approach.

\subsection{FitzHugh-Nagumo Model}

\begin{figure}[h]
    \centering
\includegraphics[scale=0.25]{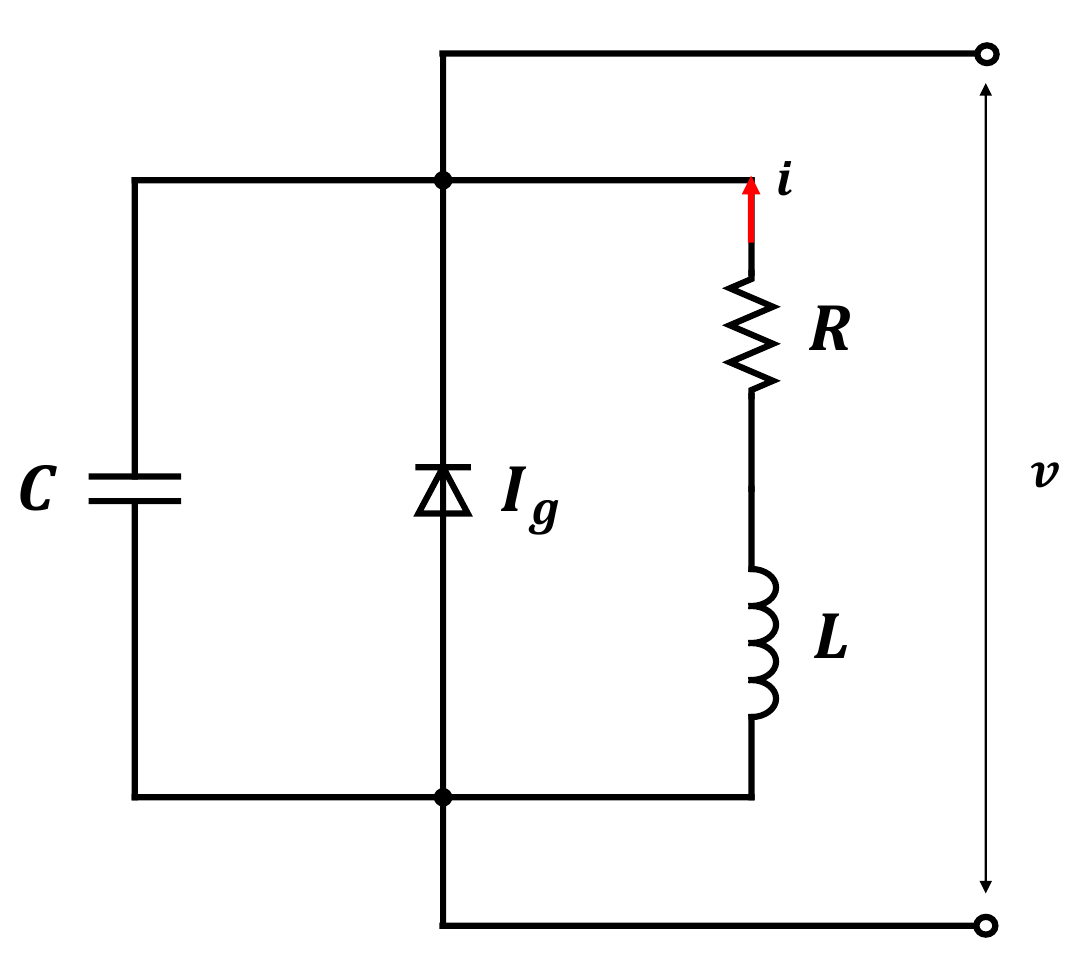}
    \caption{The FitzHugh-Nagumo model circuit is made of three branches; a linear capacitor, a linear inductor in series with a linear resistor, and finally, a tunnel diode that is a static nonlinear resistor}
    \label{fig:FN}
\end{figure}

The FitzHugh-Nagumo model captures the rhythmic behavior of a spiking neuron. The circuit representation of the model is discussed in \cite{nagumo1962active} and is shown in Fig. \ref{fig:FN}.

Using KVL and KCL, the governing equations can be obtained as
\begin{equation}
    \begin{aligned}
        C \frac{d}{dt}{v} = I_g(v) - i
\\
L \frac{d}{dt}{i} = -Ri + v
    \end{aligned}
\end{equation}
where $I_g(v) = v - \frac{v^3}{3}$. This can be rewritten as
\begin{equation}{\label{eq:FNSR}}
\begin{bmatrix}
C\operatorname{D} & +\mathbb{I}\\ 
-\mathbb{I} & L\operatorname{D}
\end{bmatrix} \begin{bmatrix}
v(t)\\ 
i(t)
\end{bmatrix}- 
\begin{bmatrix}
\operatorname{I}_g(v(t)))\\ 
-Ri(t)
\end{bmatrix} = 0
\end{equation}
where $\operatorname{D}$, again, defines the differentiation operator and $\mathbb{I}$ is the identity operator. It is possible to recast Eq. (\ref{eq:FNSR}) to the form of Eq. (\ref{eq:split}) as
\begin{equation}{\label{eq:FNDMDR}}
    \begin{bmatrix}
C\operatorname{D} & +\mathbb{I}\\ 
-\mathbb{I} & L\operatorname{D}
\end{bmatrix} \begin{bmatrix}
v(t)\\ 
i(t)
\end{bmatrix}+ 
\begin{bmatrix}
\frac{(\cdot)^3}{3} & 0\\ 
0 & R
\end{bmatrix} \begin{bmatrix}
v(t)\\ 
i(t)
\end{bmatrix} - 
\begin{bmatrix}
\mathbb{I} & 0\\ 
0 & 0
\end{bmatrix} \begin{bmatrix}
v(t)\\ 
i(t)
\end{bmatrix} = 0
\end{equation}

In the splitting of Eq. (\ref{eq:FNDMDR}), the first operator is a lossless operator that incorporates the capacitor, inductor, and interconnection of elements. The second operator represents the dissipative elements of the circuit. The last operator represents the active element that destabilizes the motion and initiates spiking. This breaks the resistive part into the difference of two monotone operators. 

\begin{figure}[h]
    \centering
\includegraphics[scale=0.20]{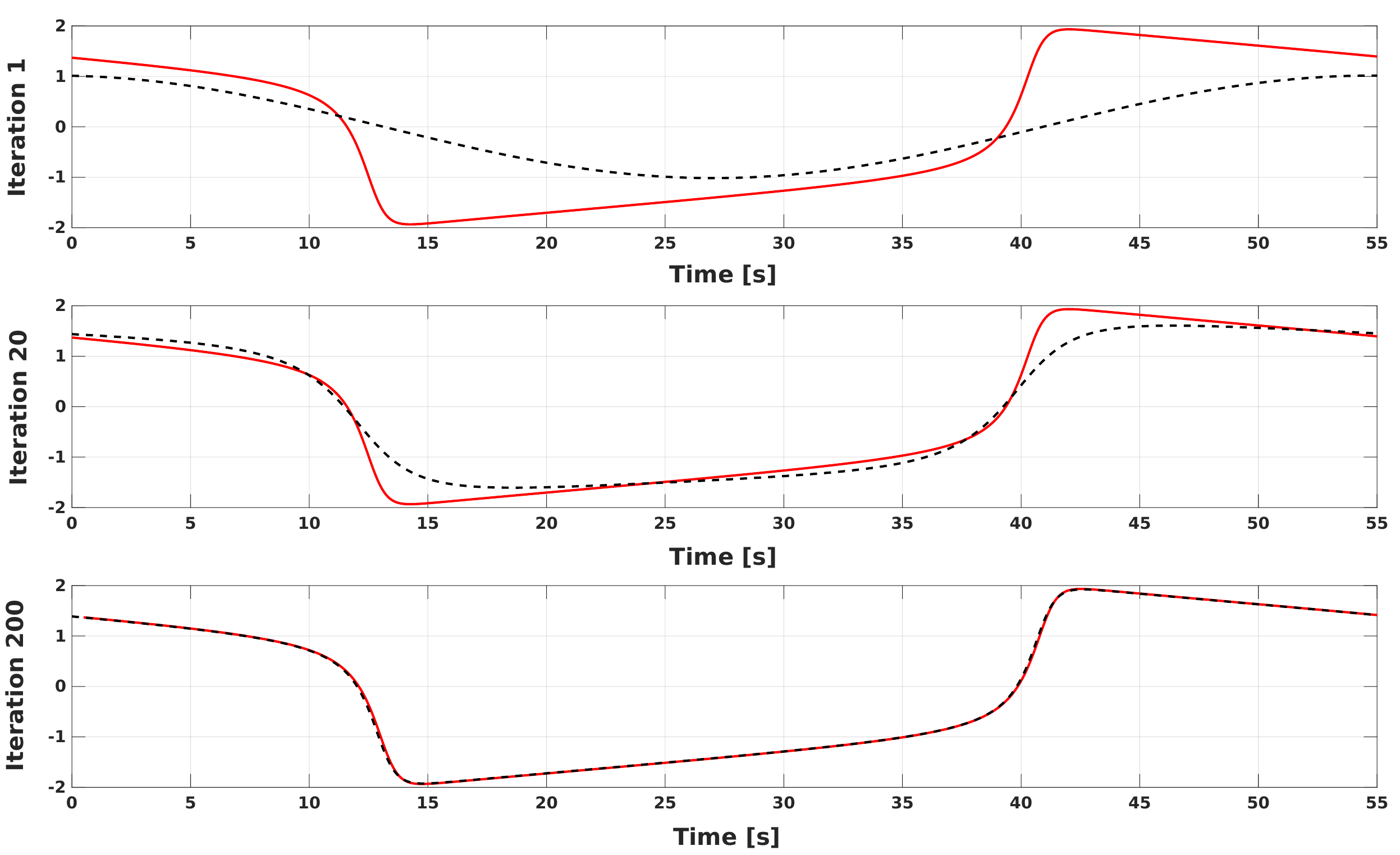}
    \caption{Simulation of the FitzHugh-Nagumo model; The result of the proposed method at iterations 1, 20, and 200 is contrasted with the steady-state results of numerical integration. The solid red line indicates the result of the settled numerical integration method (only steady-state) whereas the black dashed line is the result of the FPI method.}
    \label{fig:FPIFN}
\end{figure}

With this representation, it is straightforward to apply DMDR and obtain the solution for this circuit. Figure \ref{fig:FPIFN} illustrates the steady-state response of the FitzHugh-Nagumo circuit in the space of discrete $T$-periodic square-integrable signals (${l}^2 _T$) with parameters $L = 20$, $C = R = 1$. To proceed with the computations, the signals $v(t)$ and $i(t)$ were both discretized to vectors with 556 elements. The step size ($\alpha$) for the DMDR algorithm was set to 0.1 and the sampling frequency of the FFT was chosen to be 10 Hz. The simulation takes around 28 milliseconds to converge with a sinusoidal initial condition. The numerical integration method (two-step Adams-Bashforth method with a step size of 0.01 seconds) computes the steady-state behavior in 5 milliseconds.
\subsection{A Spiking Neural Network}
The next example studies a network of 100 neurons modeled by the FitzHugh-Nagumo circuit. The network has all-to-all connectivity meaning that each neuron is connected to every other neuron. These synaptic connections are modeled using diffusive coupling, meaning that the neurons are connected via a linear resistor. Using the resistance extraction method, the dynamics of the network can be obtained and by reformulating the dynamics of the network to the format of Eq. (\ref{eq:split}), we have
\begin{equation}
    \operatorname{S}= \left[\begin{array}{ccc:ccc}
C_1 \operatorname{D} &   & \operatorname{0}_{{l}^2 _T} & +\mathbb{I} &  & \\ 
 & \ddots   &  &  & \ddots & \\ 
\operatorname{0}_{{l}^2 _T} &  & C_{100} \operatorname{D}  &  &  & +\mathbb{I}\\  \hdashline
-\mathbb{I} &   & & L_1 \operatorname{D} &  & \operatorname{0}_{{l}^2 _T}\\ 
 &  \ddots &  &  & \ddots & \\ 
 &  & -\mathbb{I} & \operatorname{0}_{{l}^2 _T}  &  & L_{100} \operatorname{D}
\end{array}\right]
\end{equation}
and
\begin{equation}
    \operatorname{M_1}= \left[\begin{array}{ccc:ccc}
\ddots &   & \operatorname{0}_{{l}^2 _T} &\\ 
 & \frac{(\cdot)^3}{3} + \sum_{j=1, j\neq k}^{100} \frac{1}{R_{c_{k,j}}} \mathbb{I}   &  &  \operatorname{0}_{{l}^2 _T} \\ 
\operatorname{0}_{{l}^2 _T} &  & \ddots  &  \\  \hdashline
 & \operatorname{0}_{{l}^2 _T} &  & R_{k-100} \mathbb{I} \\ 
\end{array}\right]
\end{equation}
where here $k$ indicates the row number and $j$ indicates the column number. The operator in the fourth quadrant of the operator $\operatorname{M}_1$ is diagonal, meaning $R_{k-100} \mathbb{I}$ is on the main diagonal of this matrix and the rest of the elements are the zero operator. Also
\begin{equation}
    \operatorname{M_2}= \left[\begin{array}{cccc:ccc}
\mathbb{I} &  \frac{1}{R_{c_{1,2}}}\mathbb{I} & \ldots &  \frac{1}{R_{c_{1,100}}}\mathbb{I} & &  & \\ 
\frac{1}{R_{c_{2,1}}}\mathbb{I} & \mathbb{I}   &  & \vdots &    &  &\\ 
\vdots &    & \ddots &    &  & \operatorname{0}_{{l}^2 _T} &\\ 
\frac{1}{R_{c_{100,1}}}\mathbb{I}  &  \ldots  &    & \mathbb{I} &  &  &\\ 
 \hdashline
 &    &  &  & &  &\\ 
 &   \operatorname{0}_{{l}^2 _T} &   &  & & \operatorname{0}_{{l}^2 _T} &\\ 
 &   &   &  & &  &
\end{array}\right]
\end{equation}
where $\operatorname{0}_{{l}^2 _T}$ is the zero operator of adequate dimension, and $L_j$, $C_j$ and $R_j$ are the inductance, capacitance and resistance of the $j^{\text{th}}$ neuron. All the four blocks of operators $\operatorname{S}$, $\operatorname{M_1}$, and $\operatorname{M_2}$ are of the same dimension. $R_{c_{i,j}}$ represents the strength of the diffusive coupling between neurons $i$ and $j$ and thus $R_{c_{i,j}} = R_{c_{j,i}}$. The nominal values of these parameters are $C_i = 1$, $L_i = 20$, $R_i = 1$ and $R_{C_i} = 5$. Nevertheless, to incorporate heterogeneity into the simulation, a random deviation, up to 20 \% of the nominal values, is added to all the parameters.

Simulating this network, the algorithm converges to the true response in approximately 2.91 seconds with sinusoidal initial conditions. The numerical integration method computes the steady-state in 3.08 seconds. Compared with the example of a single neuron, the proposed method is observed to have better scalability. All the simulations are performed in an HP tower Z2. The graphical results of the network example can be accessed through the provided simulation code.


A final point that must be noted is the behavior-informed nature of this solver. From the previous example, the shape of a FitzHugh-Nagumo spike was obtained. It is also known that strong diffusive couplings lead to synchronous behavior. Now, it is possible to initialize this simulation with this knowledge (synchronized spikes where the shape of the spikes is extracted from the previous example) and allow the solver to converge. It will be then observed that the solver will converge extraordinarily fast (28 milliseconds) since the initial guess is close to its solution. If it had not been for the introduced heterogeneity, the solver would have converged instantaneously. This is a {\textit{behavior-informed}} simulation.

\section{Conclusions}
Nonlinear spiking neural networks are simulated by fixed-point iteration methods of large-scale convex optimization. To this end, an energy-based splitting is proposed that breaks the governing operator of the nonlinear network into an LTI lossless part and a nonlinear resistive part. The nonlinear resistive part is further broken into a difference of two cyclic monotone operators. Fixed point iteration methods that deal with difference of monotone systems are introduced and used to solve these networks. It is also shown that by switching between the time domain and the frequency domain, the LTI structure of the lossless operator can be exploited and the computations can become efficient. A large-scale network of spiking neurons modeled by the Fitzhugh-Nagumo circuit and diffusive coupling is simulated using this method to demonstrate its computational tractability and scalability.

\section{Code Availability}
All the code corresponding to the methods and simulations of this paper is publicly available at \href{https://github.com/AmirShahhosseini/FitzHugh_Nagumo_CDC}{this} link. 

\section{Future Works}
Extending the method of this paper to richer models of neurons that capture complicated behavior such as {\textit{bursting}} is necessary. Additionally, the modeling of active synaptic connections sophisticates the entire process but is necessary for simulating and analyzing neuromorphic and neuroscientific systems. Finally, the DMDR method is impractical for decompositions that split the problem into more than three operators. This algorithm must be extended to allow for the analysis of more complex neuron models and synaptic connections.
\bibliographystyle{IEEEtran}
\bibliography{ref}

\end{document}